\documentclass[fleqn,usenatbib]{mnras}

\usepackage{newtxtext,newtxmath}

\usepackage[T1]{fontenc}

\DeclareRobustCommand{\VAN}[3]{#2}
\let\VANthebibliography\thebibliography
\def\thebibliography{\DeclareRobustCommand{\VAN}[3]{##3}\VANthebibliography}

\usepackage{graphicx}	
\usepackage{amsmath}	
\usepackage[table]{xcolor}
\usepackage{multirow}
\usepackage{pifont}

\newcounter{qnumber}

\title[99 Her Polar Planets]{The polar debris disc around 99 Herculis: A potential signpost for polar circumbinary planets}

\author[J. L. Smallwood et al.]{
Jeremy L. Smallwood,$^{1}$\thanks{E-mail: drjeremysmallwood@gmail.com} William DeRocco,$^{2,3}$ Zhizhen Qin,$^{4}$
Antranik A. Sefilian$^{5}$ 
\\
$^{1}$Homer L. Dodge Department of Physics and Astronomy, University of Oklahoma, 440 W. Brooks St., Norman, OK 73019, USA \\
$^{2}$Maryland Center for Fundamental Physics, University of Maryland, College Park, 4296 Stadium Drive, College Park, MD 20742, USA\\
$^{3}$Department of Physics \& Astronomy, The Johns Hopkins University, 3400 N. Charles Street, Baltimore, MD 21218, USA\\
$^{4}$Department of Earth Sciences, The University of Hong Kong, PokFuLam, Hong Kong
\\
$^{5}$Department of Astronomy and Steward Observatory, The University of Arizona, 933 North Cherry Ave, Tucson, AZ, 85721, USA 
}

\date{Accepted XXX. Received YYY; in original form ZZZ}

\pubyear{2025}

\begin{document}
\label{firstpage}
\pagerange{\pageref{firstpage}--\pageref{lastpage}}
\maketitle

\begin{abstract}
The nearby binary star system 99 Herculis (99 Her) is host to the only known polar-aligned circumbinary debris disc. We investigate the hypothesis that the narrow structure of this circumbinary disc is sculpted by the gravitational influence of one or more unseen polar circumbinary planets. We first establish the theoretically viable parameter space for a sculpting planet by considering dynamical stability and clearing mechanisms, including the chaotic zone, Hill radius, diffusion,  and polar alignment timescales. We then use $N$-body simulations to test three specific architectures: a single planet interior to the disc, a single planet exterior, and a two-planet system bracketing the disc. Our simulations demonstrate that single-planet models are insufficient to reproduce the observed morphology, as they can only truncate one edge of the disc while leaving the other dynamically extended. In contrast, the two-planet shepherding model successfully carves both the inner and outer edges, confining the debris into a narrow, stable polar ring consistent with observations. We conclude that the structure of the 99 Her debris disc is most plausibly explained by the presence of two shepherding, polar circumbinary planets. We present a specific, testable model for this unique system, which elucidates the pivotal role of planetary bodies in sculpting the architecture of debris discs.
\end{abstract}

\begin{keywords}
binaries: 99 Herculis -- planet–disc interactions -- planets and satellites: dynamical evolution and stability
\end{keywords}


\section{Introduction}
\label{sec:intro}

Due to the inherent short-period biases of radial-velocity and transit surveys and the limited sensitivity of present direct imaging surveys, wide-separation planets remain a challenging target for detection. However, the presence of wide-orbit planets can often be indirectly inferred by their influence on observable structures within their system, such as debris discs \citep{Wyatt1999, Krivov2010, Hughes+2018}. Debris discs are thought to be shaped by their interactions with nearby planetary companions, serving as signposts of their presence and providing insight into their masses and orbital properties \citep{kennedy2020,Pearce2022}. Debris discs appear to be fairly ubiquitous outcomes of late-time system evolution, and have been observed in a wide array of both single-star \citep{Matra2025} and binary systems \citep{Trilling2007, Yelverton2019}. 

Though many debris discs have been observed around binary systems in coplanar or slightly misaligned configurations, both theory and simulations indicate that such debris discs should also be found in polar circumbinary configurations, with their orbital plane roughly perpendicular to the orbital plane of the binary system \citep{Kennedy2012,Smallwood2020,Smallwood2024b}. This seemingly exotic configuration is in fact a stable equilibrium for circumbinary orbits and is expected to arise naturally when a circumbinary protoplanetary disc is initially misaligned relative to an eccentric binary \citep{Aly2015,Martin2017,Zanazzi2018,Johnson2025}.
Misaligned circumbinary discs can evolve rapidly (on $\mathrm{kyr\!-\!Myr}$ timescales), causing the planetary system to settle into a polar configuration in the course of gas dispersal \citep{Smallwood2020,Smallwood2024b}. At later times, any polar circumbinary planets that have formed can gravitationally shape the leftover polar debris disc.

However, despite the theoretical motivation for the existence of such systems, observations remain scarce. At present, no polar circumbinary planets have been directly observed -- though their presence has been suggested in systems such as AC Her \citep{Martin2023} and 2M1510 \citep{Baycroft2025,Smallwood2025b} -- and only one polar debris disc has been observed. This debris disc lies in the 99 Herculis (99 Her) system, a 6 -- 10 Gyr-old binary roughly 15 pc from Earth \citep{Kennedy2012}. The presence of this disc makes the 99 Her system a unique opportunity to not only test theoretical models of protoplanetary disc evolution in binary systems \citep{Martin2017} and polar debris disc dynamics \citep{Wang2023}, but also to potentially characterize the properties of yet-unseen polar circumbinary planets within the system. In this Letter, we analyze the polar-aligned debris disc around 99 Her and investigate the potential for such a disc to be sculpted by associated polar circumbinary planets. As we show below, we find that the observed properties of the debris disc are best explained by the sculpting of two nearby polar circumbinary planets and infer the potential orbits and masses of such worlds.

This Letter is organized as follows: Section~\ref{sec:99her} details the observed morphology of the debris ring as characterized by \citet{Kennedy2012}. Section~\ref{sec:constraints} outlines the dynamical constraints on planet mass and semi-major axis required to sculpt the disc. Section~\ref{sec::nbody} presents the results of $N$-body simulations of planets interacting with a polar-aligned disc. Finally, we discuss our conclusions in Section~\ref{sec:conc}.

\section{The 99 Her Binary system and Polar Debris disc}
\label{sec:99her}
99 Her is an exemplary system for studying debris disc dynamics in binary environments due to its one-of-a-kind polar debris disc and well-constrained binary orbit. Historical astrometric observations date back to the late 1800s \citep[e.g.,][]{Burnham1878,Flammarion1879,Gore1890}, establishing a long observational baseline. The system comprises an F7V primary and a K4V secondary, with the primary exhibiting an estimated age of 6–10 Gyr, consistent with main-sequence evolutionary models \citep{Nordstrom2004,Takeda2007}.

Recent high-precision astrometry has further refined the binary orbital parameters. \cite{Kennedy2012} derived these parameters by fitting position angles (PAs) and separations from the Washington Double Star (WDS) Catalog \citep{Mason2018}. The orbital solution yields a semimajor axis of $a_{\rm b} = 16.5\, \rm au$, eccentricity $e_{\rm b} = 0.766$, inclination $i_{\rm b} = 39^\circ$, orbital period $P_{\rm orb} = 56.3\, \rm yr$, longitude of the ascending node $\Omega = 41^\circ$, argument of pericentre $\omega = 116^\circ$, and position angle $\rm PA = 163^\circ$. The apparent discrepancy between $\Omega + \omega$ and the sky plane position angle is a consequence of the system's orbital inclination.  Using a distance of $15.64\, \rm pc$ \citep{vanLeeuwen2008}, the total system mass is inferred to be $M_{\star} = 1.4\, \rm M_{\odot}$. The spectroscopic mass function further implies a mass ratio of 0.49, leading to individual masses of $M_1 = 0.94\, \rm M_{\odot}$ for the primary and $M_2 = 0.46\, \rm M_{\odot}$ for the secondary.

The debris disc of 99 Her was first detected in the 100 µm and 160 µm bands by the Herschel Photodetector and Array Camera and Spectrometer \cite[PACS;][]{Poglitsch2010}, with subsequent unresolved detections at $250$ and $350\, \rm \mu m$ by a Spectral and Photometric Imaging Receiver \cite[SPIRE;][]{Griffin2010}. Analysis of resolved PACS imaging using a two-dimensional (2D) Gaussian model by \citet{Kennedy2012} yielded the disc's properties, where the model's width, axis ratio, and major axis orientation determine the apparent size, inclination, and position angle, respectively. While this method effectively measures the apparent shape of the light emission, it does not represent a detailed physical model of the disc's internal structure. Their best-fit model is a debris ring confined to a narrow radial zone near $120\pm 10\, \rm au$, although the grain properties and size distribution remain unconstrained due to the limited spectral coverage.

The projection of the binary’s pericentre onto the sky yields a PA of $163^\circ \pm 2^\circ$, with the perpendicular direction at $73^\circ \pm 2^\circ$. Given that the debris disc's observed PA is $72^\circ$, \cite{Kennedy2012} inferred that the disc is oriented at approximately $87^\circ$ relative to the binary pericentre direction -- merely $3^\circ$ offset from a strictly polar configuration. They note, however, that a mirror inversion of the ring on the sky plane could alternatively imply a $30^\circ$ misalignment. To discriminate between these scenarios, circumbinary test particle simulations were conducted. These simulations, which account for secular perturbations due to the binary demonstrate that a $30^\circ$ misaligned debris ring would diffuse into a broader structure over time. Assuming the largest fragments are on the order of 1 mm, the secular precession period is estimated at approximately 0.5 Myr \citep{Kennedy2012}, leading to the dissolution of the ring structure within roughly 5 Myr (ten precession cycles). Thus, the nearly polar configuration is the most consistent with the observed PACS imagery. In what follows, we will use the disc orbital parameters and width inferred by \citet{Kennedy2012} to place constraints on potential polar circumbinary planetary companions, taking them as truth values. However, we note that if future high-resolution imaging of the system changes or refines the preferred disc model, our constraints would change accordingly.


\section{Dynamical Constraints on Planetary Orbits}
\label{sec:constraints}


There are a variety of dynamical arguments that can be used to place constraints on the potential orbital properties of any polar circumbinary planets in this system. In the following, we assess four primary constraints: (1) that the debris disc does not lie within the chaotic region of a planet's orbit, (2) that its edge can be sculpted by a nearby planet, (3) that it can form within the system's age and be stable, and (4) that the inferred planet(s) are most likely to have a polar orbit. We also consider the dynamically unstable region due to the perturbations from the central binary, which extends out to $\sim 1.5 a_{\rm b}$ \citep{Kennedy2019}.

\subsection{Chaotic region}
The overlapping of a planetary body's orbital resonances causes dynamical chaos that is responsible for orbital instabilities \citep{Lecar1992,Holman1993,Lecar2001,Tsiganis2005}.
For the simplest case, the restricted three-body problem, the width of the chaotic zone near a planet is the region where the first-order mean-motion resonances are wide enough to overlap \citep{wisdom1980}. A test particle that is located within the chaotic region will be strongly perturbed in a chaotic manner; as such, the chaotic zone can be used to estimate the masses of unseen perturbers that may be responsible for clearing observed gaps and truncating disc edges \citep{Quillen2006,Chiang2009,Moro-mart2010,MustillWyatt2012,Schneider2014,Rodigas2014}. 

\subsubsection{Circular planets}

The half-width of the chaotic zone on either side of a circular planet is:
\begin{equation}
    \Delta a = c \mu^{2/7}a_{\rm p},
    \label{eq::chaotic}
\end{equation}
where $\mu$ is the planet-star mass ratio, $a_{\rm p}$ is the semi-major axis of the planet, and $c$ is a numerical coefficient.  In practice, the numerical coefficient is set to $c = 1.3$ for a  planet on a circular orbit \citep{Morrison2015}. However,  an alternate analytical study by  \citet{Malhotra1999}  derived $c=1.4$ and a numerical study by \citet{Duncan1989} measured $c\approx1.5$. Although Equation~(\ref{eq::chaotic}) is derived using a circular orbit planet, the numerical coefficient $c = 1.3$ has been shown to be valid for planet eccentricities up to $e_{\rm p} \sim 0.3$ \citep{Quillen2006}.  Thus, in what follows, we set $c = 1.3$.

\begin{figure} \centering
 \includegraphics[width=0.90\columnwidth]{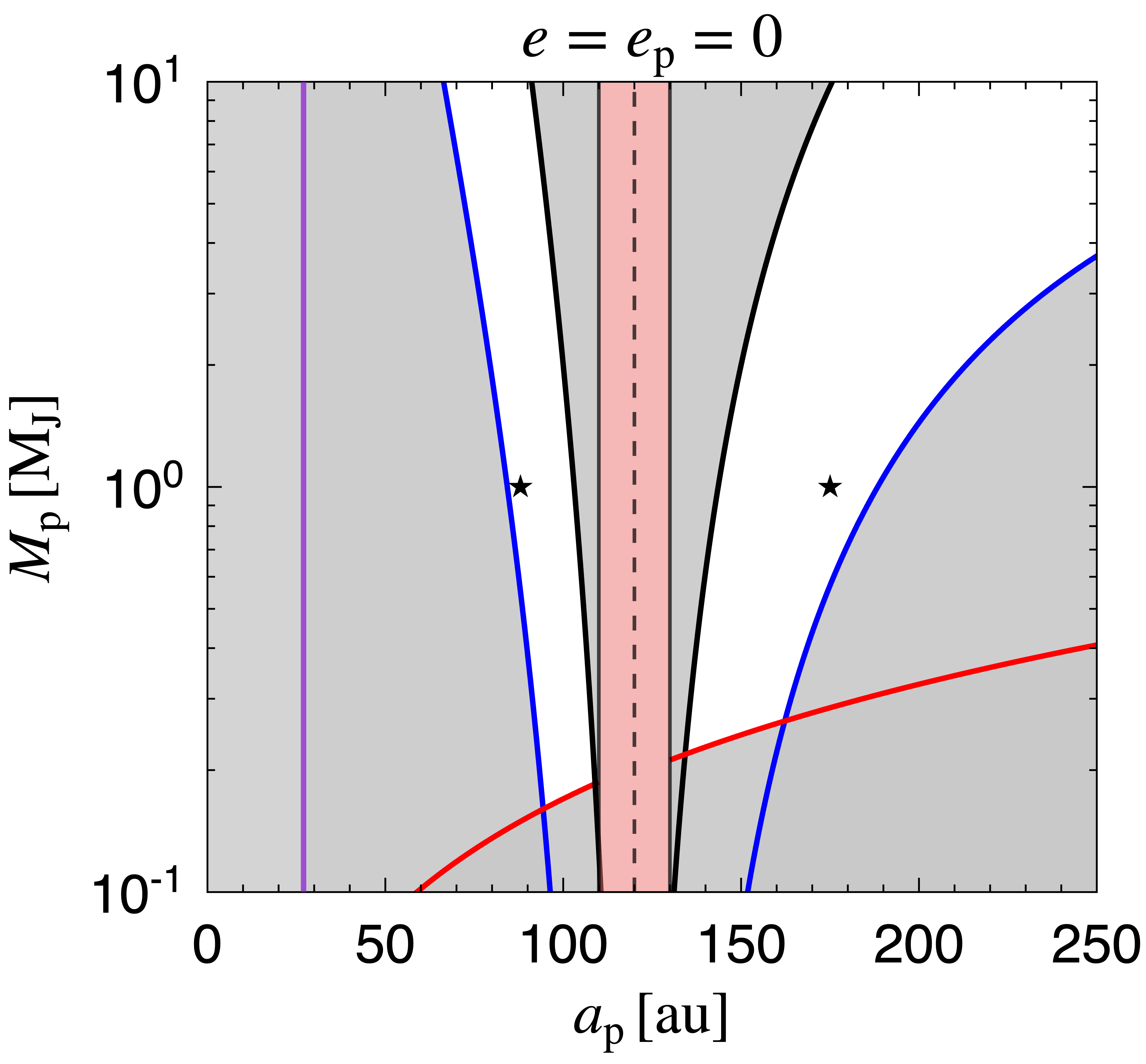}
\caption{Dynamical constraints on the semi-major axis ($a_{\rm p}$) and mass ($M_{\rm p}$) of a hypothetical polar circumbinary planet in the 99 Her system for eccentricity $e = e_{\rm p} = 0$. The observed location of the debris disc is marked by the vertical red band at $120\pm 10\, \rm au$. The purple line denotes the dynamically unstable region due to perturbations from the central binary   ($a_{\rm p} \lesssim 27\, {\rm au}\, (1.5a_{\rm p})$). The lines represent constraints from the chaotic zone (black; Eq.~\ref{eq::M_chaos}), Hill radius sculpting (blue; Eq.~\ref{eq::hill_rad}), and diffusion timescale (red; Eq.~\ref{eq::diff_time}). The unshaded (white) region indicates the viable parameter space for a planet capable of sculpting the inner and/or outer edge of the observed polar debris disc. Star symbols ($\star$) mark the parameters chosen for the $N$-body simulations presented in Section~\ref{sec::nbody}. }
\label{fig::a_M}
\end{figure}

We estimate the planetary mass required for $a_{\rm p} \pm \Delta a = Q_i$, as follows:
\begin{equation}
    M_{\rm p} \approx 1470\, M_{\rm Jup} \bigg( \frac{M_\star}{1.4 M_\odot} \bigg) \bigg[  \bigg \lvert \frac{Q_{\rm i}}{a_{\rm p}} -1  \bigg \rvert \bigg]^{7/2} c^{-7/2},
    \label{eq::M_chaos}
\end{equation}
where $Q_{\rm i}$ corresponds to the semi-major axis of the disc edge. The absolute value quantity arises to account for $Q_i < a_{\rm p}$ and $Q_i > a_{\rm p}$. Equation~\ref{eq::M_chaos} is represented by the black curves in Fig.~\ref{fig::a_M}, which show the dynamical constraints on a planet's semi-major axis and mass needed to sculpt the 99 Her polar debris disc. We assume $Q_{\rm i} \approx 120\, \rm au$ (the observed location of the 99 Her debris ring). If a planet is located within the shaded regions bounded by the black curves, a planet's chaotic zone will disrupt the disc.



\subsubsection{Eccentric planets}

\begin{figure} \centering
\includegraphics[width=0.90\columnwidth]{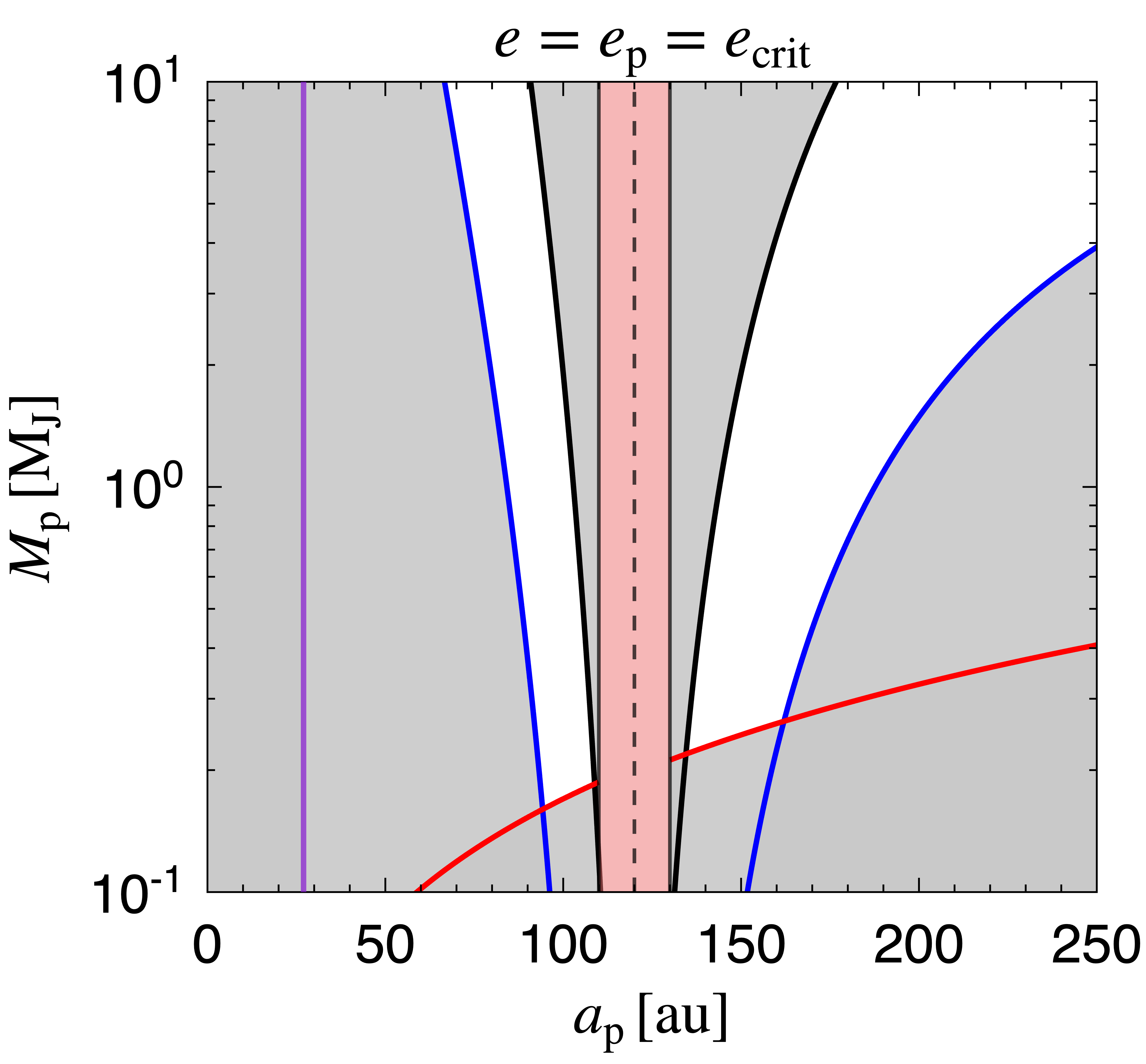}
\includegraphics[width=0.90\columnwidth]{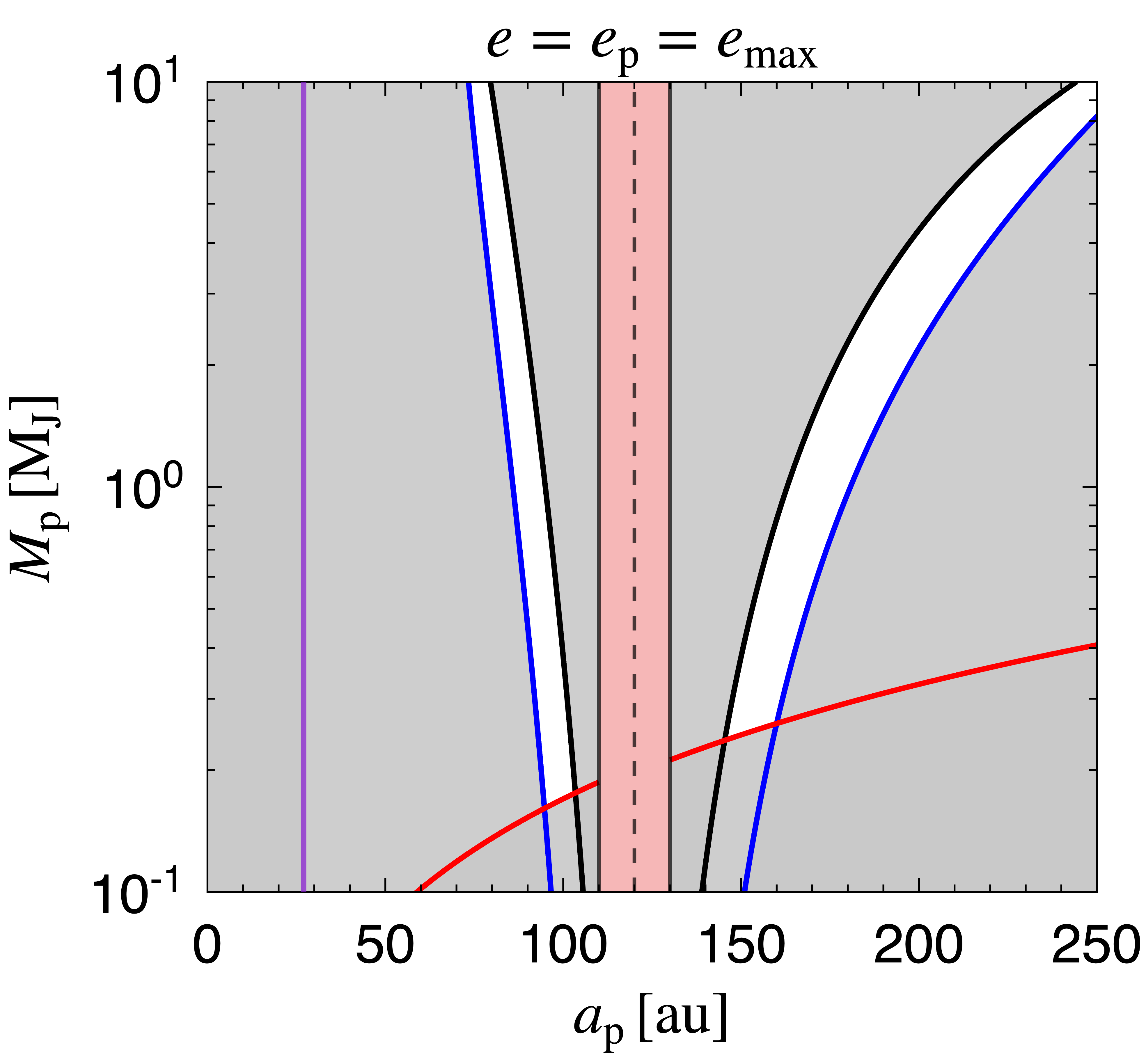}
\caption{ Same as Fig.~\ref{fig::a_M} but for eccentricities $e = e_{\rm p} = e_{\rm crit}$ (Eq.~(\ref{eq::e_crit}); top panel) and $e = e_{\rm p} = e_{\rm max}$ (Eq.~(\ref{eq::e_max}); bottom panel).  The lines represent constraints from the eccentric chaotic zone (black; Eq.~(\ref{eq::M_chaos_ecc})), eccentric Hill radius sculpting (blue; Eq.~(\ref{eq::hill_rad})), and diffusion timescale (red; Eq.~(\ref{eq::diff_time})). The unshaded (white) region indicates the viable parameter space for a planet capable of sculpting the observed polar debris disc.}
\label{fig::a_M_ecc}
\end{figure}

The half-width of the chaotic zone will be altered when the planetary and/or planetesimals orbits are sufficiently eccentric. From \cite{MustillWyatt2012}, Equation~\ref{eq::chaotic} can be rewritten to
\begin{equation}
    \Delta a_{\rm ecc} = 1.8e^{1/5}\mu^{1/5}a_{\rm p},
    \label{eq::chaotic_ecc}
\end{equation}
where $e$ is the planetesimal eccentricity. For simplicity we assume the particle eccentricity is equal to the planet eccentricity, taking $e = e_p$.  The critical particle i.e. planetesimal eccentricity that bridges the regime between Eq.~\ref{eq::chaotic} and Eq.~\ref{eq::chaotic_ecc} is given by
\begin{equation}
    e_{\rm crit} \approx 0.21\mu^{3/7}.
    \label{eq::e_crit}
\end{equation}
There is, however, a maximum eccentricity above which the extended chaotic‐zone approximation ceases to apply. At higher eccentricities, orbits enter a regime in which particles are removed primarily by direct close encounters rather than through chaotic diffusion. Consequently, the derivation for the extended chaotic zone is no longer valid beyond this threshold. The maximum eccentricity is given by
\begin{equation}
    e_{\rm max} = 2.1\mu^{1/4}.
    \label{eq::e_max}
\end{equation}
We estimate the planetary mass required for the half-width of its eccentric chaotic zone on either side to extend out to a specified radius, $Q_{\rm i}$, as follows:
\begin{equation}
    M_{\rm p} \approx 77.8\, M_{\rm Jup} \bigg( \frac{M_\star}{1.4M_\odot} \bigg) \bigg( \frac{1}{e} \bigg)\bigg[ \bigg \lvert \frac{Q_{\rm i}}{a_{\rm p}} -1  \bigg \rvert \bigg]^{5}.
    \label{eq::M_chaos_ecc}
\end{equation}
Figure~\ref{fig::a_M_ecc} reproduces the analysis from Fig.~\ref{fig::a_M}, but for planets with varying eccentricity: $e = e_{\rm p} =  e_{\rm crit}$ (top panel) and $e = e_{\rm p} =  e_{\rm max}$ (bottom panel). A more eccentric planet restricts the viable parameter space for sculpting the debris disc. 


In closing, we note that while the above calculations assume single-planet systems and may not be directly applicable to two-planet systems \citep[see, however,][]{Hadden_Lithwick_2018}, let alone a binary star, we use them as a guiding framework: the relevant effects are captured in the $N$-body simulations described in Section~\ref{sec::nbody}.

\begin{figure} \centering
\includegraphics[width=0.90\columnwidth]{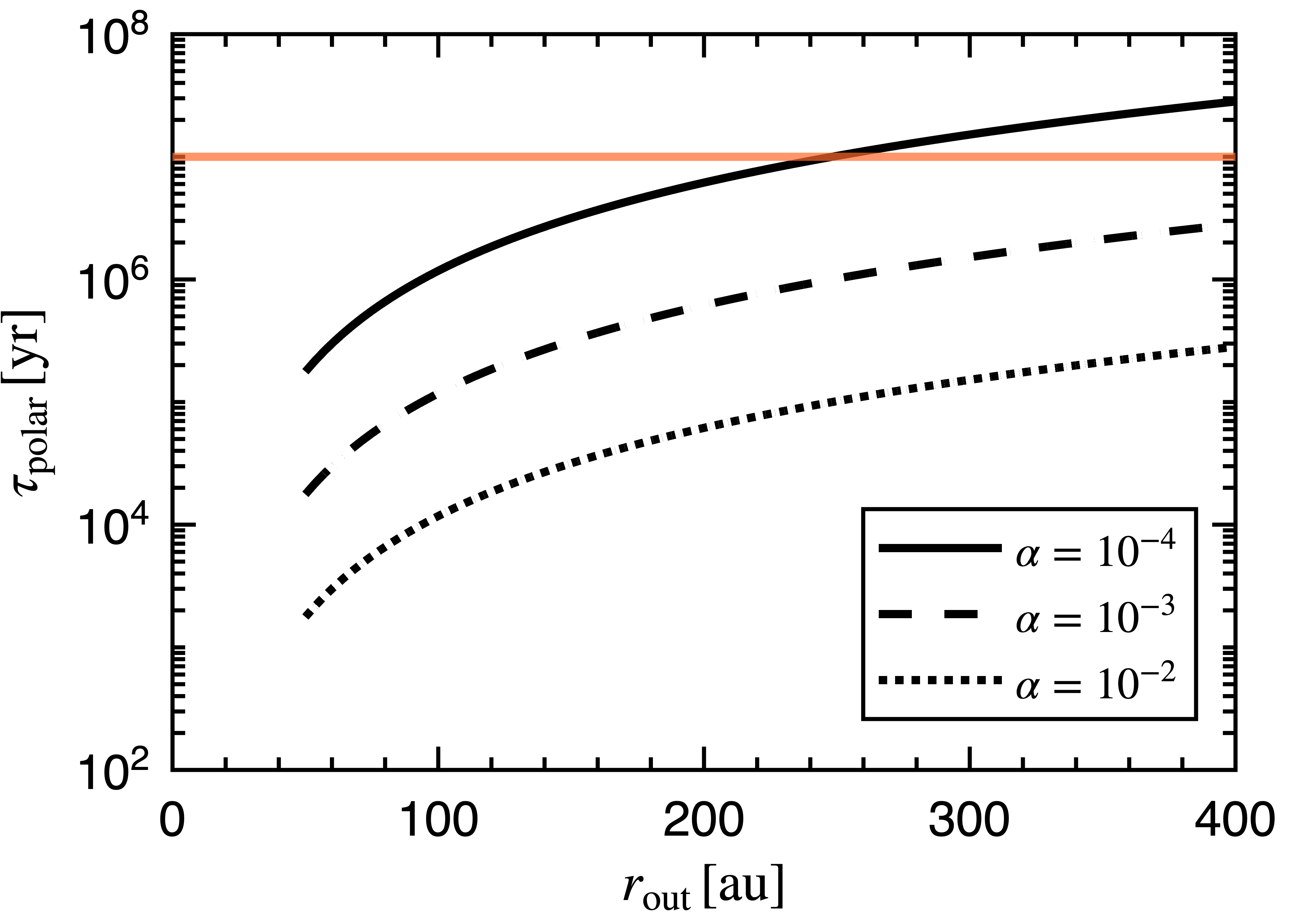}
\caption{ The polar alignment time-scale, $\tau_{\rm polar}$, as a function of outer disc radius, $r_{\rm out}$, around 99 Her for different viscosities: $\alpha = 10^{-4}$ (solid), $10^{-3}$ (dashed), and $10^{-2}$ (dotted). The orange line denotes the upper limit of the protoplanetary disc lifetime, $\sim 10^7\, \rm yr$. We find that even for low viscosity discs with $\alpha=10^{-4}$, planets inwards of $\approx 250$ au would have remained coupled to the disk during its polar alignment, lending credence to the hypothesis that planets around 99 Her are in a polar configuration.
}
\label{fig::align_time}
\end{figure}

\subsection{Hill radius}
A planet's gravitational influence can not only destabilize the debris disc, as discussed above, but help to shape its edges as well. For the specific case of clearing non-resonant debris particles from the inner parts of an external disc, we follow the methodology of \citet{Pearce2014}. They demonstrate that a planet's gravitational reach can extend to five eccentric Hill radii from the planet's apocentre \citep{Gladman1993}. Based on this, the mass of a single planet required to sculpt the disc's inner edge can be calculated as a function of $a_{\rm p}$ and $e_{\rm p}$, given by the relation:
\begin{equation}
M_{\rm p} \approx 11.73 M_{\rm Jup} \left( \frac{M_\star}{1.4 M_\odot}\right) \left[ \frac{Q_{\rm i}}{a_{\rm p}(1+e_{\rm p})} -1 \right]^3 (3-e_{\rm p}),
\label{eq::hill_rad}
\end{equation}
where $Q_{\rm i}$ is the apocentre of an ellipse that traces the inner edge of the disc, if eccentric. If we assume the disc is circular, $Q_{\rm i}$ becomes equivalent to the disc's inner edge radius. Equation~(\ref{eq::hill_rad}) can also be adapted for an outer planet truncating the outer disc edge. This relationship is illustrated by the blue curves in Fig.~\ref{fig::a_M} for a circular planet and in Fig.~\ref{fig::a_M_ecc} for an eccentric planet. We assume $Q_{\rm i}\approx 110\, \rm au$ for an interior planet and $Q_{\rm i}\approx 130\, \rm au$ for an exterior planet. These values are consistent with the disc width uncertainties from the models of \cite{Kennedy2012}. In the literature, a planet is expected to lie along the calculated curve (blue line) to clear debris from a specific location. However, given the positional uncertainty of the debris disc ($120\pm10\, \rm au$), this curve represents a lower bound on the planet's possible location 
if it lies interior to the disc, and an upper bound if it lies exterior.

\begin{figure*} \centering
\includegraphics[width=1.9\columnwidth]{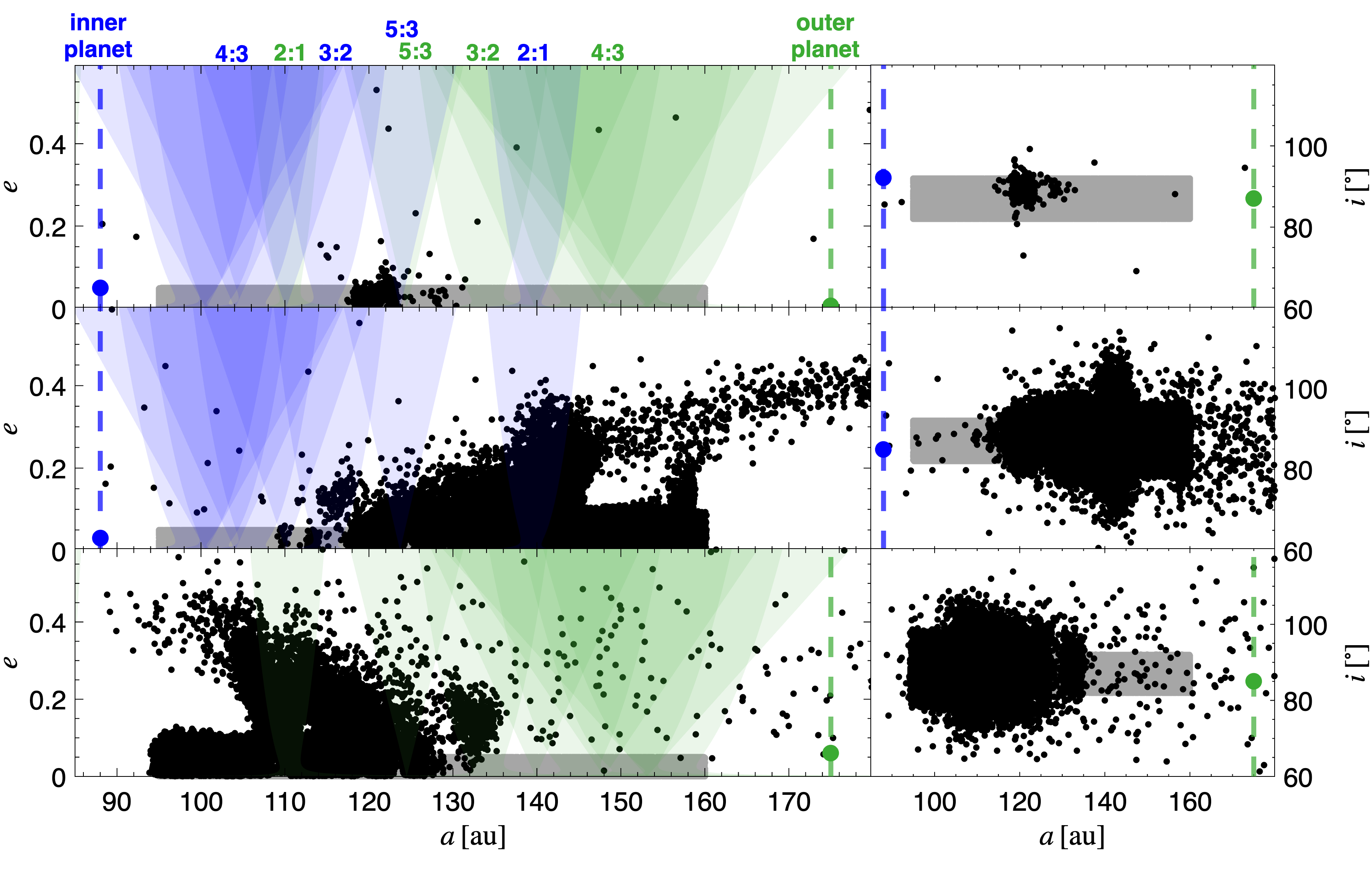}
\caption{The left-hand panels show the eccentricity $e$ versus semi-major axis $a$, while the right-hand panels show inclination $i$ versus $a$. All panels illustrate the final state of the debris disc after a 100 Myr integration. The three panels display different planetary configurations: a two-planet system (top), an inner-planet-only system (middle), and an outer-planet-only system (bottom). The initial distribution of test particles is shown in gray, while the surviving particles are shown using black circles. The vertical dashed blue and green lines mark the initial locations of the inner and outer planets, respectively, while the corresponding dots show their final semi-major axes and eccentricities or inclinations after 100 Myr.  Shaded regions indicate the locations of mean-motion resonances associated with the inner (blue) and outer (green) planets, with several prominent resonances labeled. Although we only show the widths for first and second-order resonances, the third-order 5:2 MMR visibly excites eccentricities near 160 au in the middle panel.}
\label{fig::a_e_plot}
\end{figure*}

\subsection{Diffusion timescale}

To establish a lower limit on the planet's mass, the planet must be massive enough to sculpt the disc by clearing debris within the stellar lifetime ($t_\star$). We adopt the framework of \citet{Pearce2014}, where the total clearing time is approximately 10 times the longer of two key timescales: the diffusion time for scattering nearby particles and the secular time for perturbing distant particles onto planet-crossing orbits. Based on the age of 99 Her, the diffusion timescale is the more relevant process. This allows us to calculate a minimum planet mass by inverting the formulation for the scattering-dominated regime, which is given by (see also \citealt{Tremaine1993}):
\begin{equation}
    M_{\rm p} \geq 0.426\, M_{\rm Jup} \bigg(\frac{a_{\rm p}}{\rm au}\bigg) \bigg(\frac{Q_{\rm i}}{\rm au}\bigg)^{-1/4} \bigg(\frac{t_\star}{\rm Myr}\bigg)^{-1/2} \bigg(\frac{M_\star}{1.4M_{\odot}}\bigg)^{3/4}.
    \label{eq::diff_time}
\end{equation}
Equation~\ref{eq::diff_time} is represented by the red curve in Figs.~\ref{fig::a_M} and~\ref{fig::a_M_ecc}. For 99 Her, which has minimum age of $t_\star \sim 6\, \rm Gyr$, the diffusion timescale does not significantly restrict the parameter space. Its primary effect is to require that a planet be sufficiently massive to have cleared the debris within the system's lifetime.

\subsection{Polar alignment timescale}
We expect that, since the debris disc is observed to be polar-aligned, provided that planet formation occurred after (or during the late stages of) the disc's polar alignment. The same may not necessarily be true for any planets exterior to the debris disc, given that the polar alignment timescale of the primordial gaseous disc increases with disc radius. Following the calculations of \cite{Smallwood2020}, Figure~\ref{fig::align_time} shows the polar alignment timescale $\tau_{\rm polar}$ as a function of the primordial circumbinary disc's outer radius around 99 Her with a disc aspect ratio $H/r = 0.05$. For less viscous discs, regions within an outer radius of $\sim 250\,\rm au$ can achieve polar alignment within a typical disc lifetime of $\sim 10\,\rm Myr$ (notwithstanding that circumbinary disc lifetimes may be longer \cite[e.g.,][]{Mamajek2014,Czekala2015}). In contrast, for more viscous discs, the disc can extend beyond $400\,\rm au$ and still become polar-aligned on this timescale. Since the planets investigated in our $N$--body simulations have semi-major axes less than $250\,\rm au$ (Section \ref{sec::nbody}),  we expect that the gas disc at these radii would have achieved polar alignment within the disc lifetime. However, we note that if planet formation (or significant dust growth and clumping) proceeded rapidly enough to decouple from the gas before polar alignment was complete, non-polar planets could potentially form. As such, future observations of this system could provide a test for the rate of planet formation, as the observation of polar-aligned planets would indicate that alignment outpaced planet formation in this system.
 Nevertheless, \cite{Childs2022} simulated the dynamical evolution of misaligned, gas-free debris discs and found that highly misaligned systems tend to produce planets in near-polar orbits, as these systems undergo more mergers that lead, on average, to smaller misalignment angles. A detailed exploration of the relative timescales of planet formation versus disc alignment in the 99 Her system is beyond the scope of this Letter but warrants future investigation.

\section{\texorpdfstring{$N$}{N}-body simulations}
\label{sec::nbody}

We verify the analytical constraints presented in the previous section by running dedicated $N$-body simulations of the 99 Her system to see whether planets lying within our allowed parameter space do successfully reproduce the observed structure of the debris disc. In the following, we present an orbital configuration that well-reproduces the disc. Furthermore, we show that the apparent narrow width of the disc suggests that there may be two planets in the system whose combined effects help to confine the disc to a width of $\approx 20$ AU.

\subsection{Setup}
We model the 99 Her system using the WHFast integrator, which is a second-order symplectic Wisdom-Holman integrator with 11th-order symplectic correctors implemented in the $N$-body simulation package REBOUND \citep{Rein2012}. To take advantage of WHFast's improved error handling \citep{Rein2015}, which achieves Brouwer's law \citep{Brouwer1937}, we use a timestep of 3.5 percent of the binary orbital period. The integrator employs a fast and accurate Kepler solver along with unbiased coordinate transformations between Jacobi and inertial frames, ensuring numerical stability and energy conservation over long integration timescales. We perform the simulations in the center-of-mass frame using the observed orbital parameters for 99 Her, with the binary positioned initially at apastron. Our Cartesian coordinate system is defined with the x-axis aligned along the binary’s eccentricity vector and the z-axis along its angular momentum vector.

To investigate the structure and dynamics of the debris disc in the 99 Her system, we simulate a disc composed of 100,000 test particles.
 That is, we neglect the disk’s gravity, although recent studies suggest it can be important even when the disk is less massive than external perturbers \citep{Sefilian2024, Sefilian+2025}.
Initially, the disc extends radially from $r_{\rm in} = 95\, \rm au$ to $r_{\rm out} = 160\, \rm au$. The initial orbital elements of the particles are drawn from uniform distributions: eccentricities from 0 to 0.05; inclinations from $82^\circ$ to $92^\circ$ (to model a nearly polar alignment with the binary orbital plane); and longitudes of the ascending node from $82^\circ$ to $92^\circ$ (consistent with the observed disc tilt of approximately $87^\circ$). The remaining elements, the argument of pericenter and the true anomaly, are randomized between from $-180^\circ$ to $180^\circ$. We integrate the system for $100\, \rm Myr$. Although this duration is less than the system's age, it is significantly longer than the secular timescale at $120\, \rm au$ ($t_{\rm sec} \sim 10^5\, \rm yr$; \citealp{Kennedy2012}), ensuring that secular and resonant effects are fully captured. Particles with a semi-major axis $a > 1000\, \rm au$ or an eccentricity $e \geq 1$ are considered ejected and are removed from the simulation.

We simulate three dynamical configurations to investigate the gravitational interactions of the binary--planet--debris disc system. The first scenario involves two planets bracketing the debris disc—one interior and one exterior—to study their combined effects on its structure and dynamics. The second scenario places a single planet interior to the disc to analyze its influence on the inner edge, while the final configuration positions a planet exterior to the disc to examine its effect on the outer edge. In all cases, the planets are Jupiter-mass bodies on initially circular orbits\footnote{Over the course of the simulations, interactions with the central binary and additional planets (if present) excite non-zero planetary eccentricities.}, inclined at $87^\circ$ with respect to the binary orbital plane to match the observed polar circumbinary debris disc. The inner planet's semi-major axis is set to $a_1 = 87\, \rm au$, and the outer planet's semi-major axis is set to $a_2 = 175\, \rm au$ (shown by the star symbols in Fig.~\ref{fig::a_M}). The stability of polar circumbinary planets have been discussed previously \citep{Cuello2019,Chen2020}. While not reported here, we also conducted long-term (Gyr-timescale) stability simulations for single- and multi-planet polar circumbinary configurations around 99 Her. These integrations confirm that the planetary architectures selected for our $N$-body simulations are stable over the age of the system.

\subsection{Results}
Figure~\ref{fig::a_e_plot} shows the final structure of a polar-aligned debris disc after 100 Myr of evolution under the influence of one or more polar circumbinary planets. In all three scenarios, the planets effectively sculpt the disc by clearing material from their chaotic zones (shown by the overlapping mean-motion resonances widths) and exciting the eccentricities and inclinations of remaining particles through secular and resonant interactions. The initial, dynamically cold state of the disc (shown in gray) is quickly disrupted, leading to distinct final architectures (black particles) that are dictated by the planetary configuration.

The top panels reveal the combined effect of both an inner and an outer planet. This two-planet architecture proves highly effective at confining the disc into a narrow ring. The resulting ring is located between the 3:2 and 5:3 mean-motion resonances of the inner planet and the 2:1 and 5:3 mean-motion resonances of the outer planet. The inner planet carves a sharp inner boundary, while the outer planet truncates the outer edge, preventing particles from scattering to wide orbits. The region between the two planets is dynamically stable, but the eccentricities of the surviving particles are significantly smaller compared to the single-planet cases, remaining mostly below $e\approx 0.1$. Furthermore, the debris disc maintains a near-polar inclination distribution despite the gravitational influence of two giant polar circumbinary planets. This two-planet configuration robustly produces a narrow, confined debris ring by efficiently clearing the interior and exterior regions. Although the semi-major axis range suggests a radial width of less than 10 au, the actual width, defined by the innermost pericenter and outermost apocenter, is closer to 20 au due to particle eccentricities. This resulting structure closely resembles the observations and models from \cite{Kennedy2012}.

The middle panels demonstrate the effect of a single planet orbiting interior to the debris disc. This planet carves a sharp inner edge in the disc, ejecting most particles with semi-major axes less than $\sim 110\, \rm au$. The planet's gravitational influence extends throughout the disc, exciting the eccentricities of surviving particles to values up to $e \approx 0.4$. This excitation is particularly pronounced at the locations of strong mean-motion resonances (blue shaded regions), such as the 2:1 and 3:2 MMRs, which place particles onto orbits with higher eccentricities and semi-major axes. The inner planet effectively shepherds the disc, creating a broad, stirred structure. Conversely, the bottom panel shows the impact of a single planet exterior to the disc at $a_{\rm p} \approx 175\, \rm au$. This configuration efficiently sculpts the disc's outer edge, ejecting particles beyond $\sim 130\, \rm au$. The planet's interior mean-motion resonances (green shaded regions) drive significant dynamical evolution, scattering particles inward and exciting their eccentricities. In both single-planet scenarios, the inclination distribution is more broadly scattered and less tightly confined to a near-polar orientation than in the two-planet case.

\begin{figure} \centering
\includegraphics[width=0.98\columnwidth]{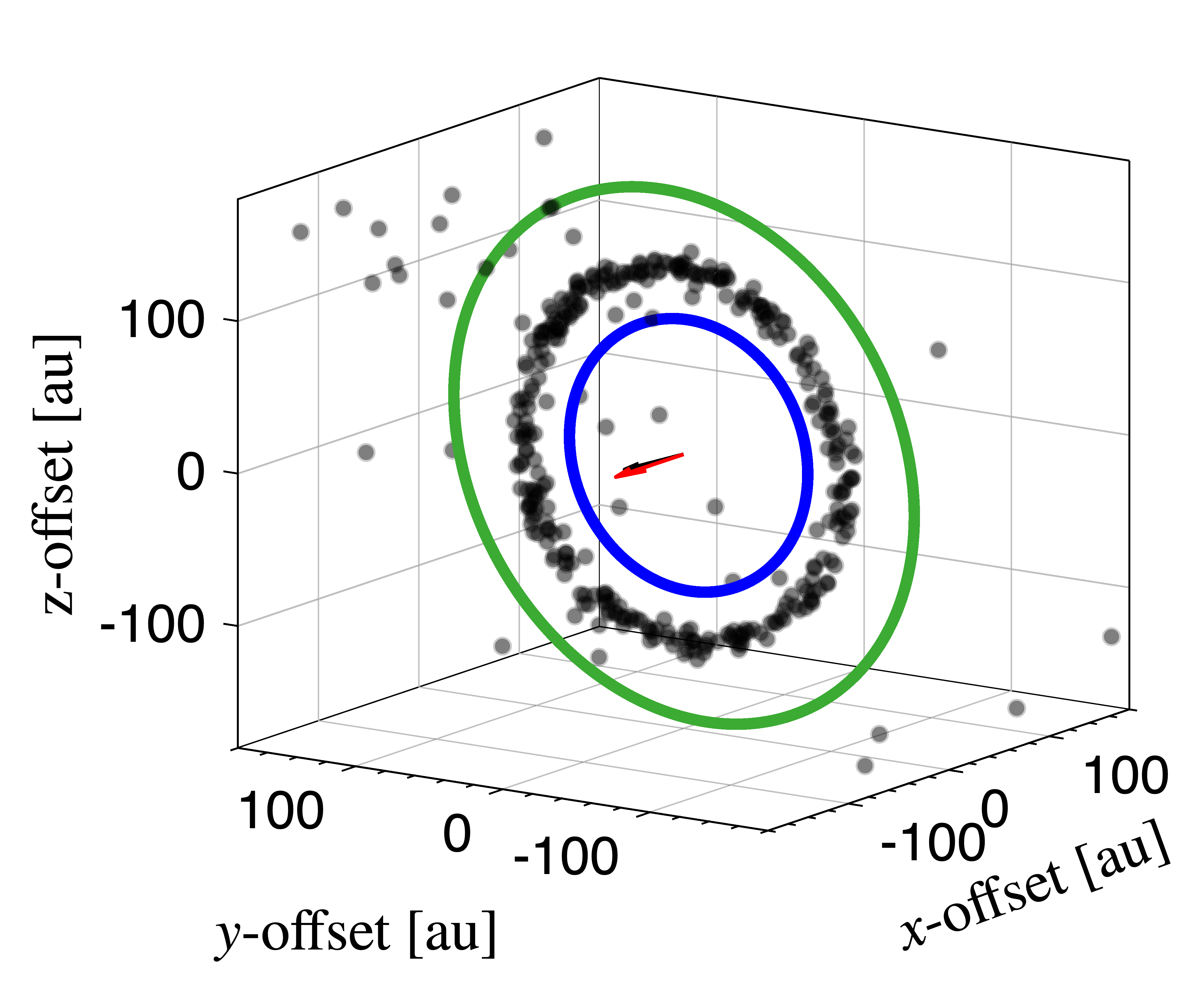}
\caption{The structure of a polar-aligned debris ring (black dots) as it is dynamically sculpted by two circumbinary planets (orbits in blue and green). The red vector denotes the binary’s eccentricity vector, while the black vector represents the net angular momentum of the particles, which is nearly aligned with the binary eccentricity vector, confirming a polar configuration. A small population of transient particles ($< 20$ per cent), which have been perturbed by the planets but not yet ejected, are excluded from the calculation of the disc's angular momentum.}
\label{fig::3d}
\end{figure}

The two-planet scenario, which brackets the debris disc, provides the best match to observations of a narrow ring centered at approximately 120 au \citep{Kennedy2012}. While single-planet models can truncate either the inner or outer edge of the disc to the desired location, they inevitably leave the opposing side extended, failing to reproduce a confined ring structure. Additionally, in either single-planet simulation, the inclination distribution is more sparse and not confined to a nearly polar orientation. In contrast, the two-planet architecture successfully sculpts the disc from both sides, and maintains the near-polar orientation. This outcome is visualized in Figure~\ref{fig::3d}, which shows the disc structure resulting from the two-planet shepherding simulation. The combined influence of the inner (blue orbit) and outer (green orbit) planets is highly effective at carving the initially broad disc into a well-defined, narrow ring. The planets efficiently clear their respective orbital regions, confining debris particles (black dots) to the zone between them. The final ring maintains a polar orientation relative to the central binary orbit, indicating that this architecture is stable.


\section{Conclusions}
\label{sec:conc}
In this work, we have investigated the hypothesis that the narrow, polar circumbinary debris disc around 99 Her is sculpted by one or more unseen polar circumbinary planets. We first established the theoretically viable parameter space for a single planet to dynamically influence the disc, considering constraints from chaotic zone and Hill radius clearing, and diffusion timescales for a circular planet (Figure~\ref{fig::a_M}) and eccentric planet (Figure~\ref{fig::a_M_ecc}). 
This analysis highlights regions where a planet could exist and interact with the disc, providing a foundation for exploring whether such interactions could produce the observed structure via $N$-body simulations.

To test this, we performed $N$-body simulations of planetary architectures chosen from the viable parameter space. Our results demonstrate that a single planet, whether interior or exterior to the disc, is insufficient to reproduce the confined morphology. While a single planet effectively truncates one edge of the disc, it invariably leaves the other side extended and dynamically excited (Figure~\ref{fig::a_e_plot}, middle and bottom panels), failing to match the observations.

In contrast, a two-planet shepherding model provides a compelling mechanism for the disc’s current structure. A system with one planet orbiting interior to the disc and a second planet exterior successfully carves both the inner and outer edges, confining the debris particles into a narrow, stable ring (Figure~\ref{fig::a_e_plot}, top panel). This architecture not only reproduces the radial confinement but also maintains the polar orientation of the disc over long timescales, as visualized in Figure~\ref{fig::3d}.

We conclude that the observed structure of the 99 Her debris disc is most plausibly explained by the presence of at least two shepherding, polar circumbinary planets. This structure is analogous to that of the Fomalhaut debris disc, which, although a single-star system, is thought to be shepherded by both an interior and an exterior planet \citep{Boley2012}. This work strengthens the case for planets as the primary architects of debris disc structures in binary systems and provides a specific, testable model for the 99 Her system.

Future, higher-resolution observations of the 99 Her system will be invaluable for testing the planet-sculpting hypothesis. The ability to resolve the disc's fine structure -- such as the precise profile of its inner and outer edges, or the presence of any gaps or asymmetries -- would provide critical diagnostics of ongoing dynamical processes. These morphological details can be directly compared with predictions from dynamical models, offering powerful new constraints on the masses and orbits of any putative planets. Ultimately, such observations are essential to determine whether the disc's remarkable polar orientation is indeed shaped and maintained by one or more polar circumbinary planets.

\section*{Acknowledgments}
The authors thank Hossam Aly for a prompt and constructive review. JLS acknowledges funding from the Dodge Family Prize Fellowship in Astrophysics. JLS is also supported by the Vice President of Research and Partnerships of the University of Oklahoma and the Data Institute for Societal Challenges.  WD acknowledges fruitful conversations with Mary Anne Limbach and Rachel Bowens-Rubin, as well as the generosity of the University of Oklahoma in hosting him during the initial stages of this project.
AAS is supported by the Heising-Simons Foundation through a 51 Pegasi b Fellowship. The computing for this project was performed at the OU Supercomputing Center for Education \& Research (OSCER) at the University of Oklahoma (OU).

\section*{Data Availability}
The data supporting the plots within this article are available on reasonable request to the corresponding author. A public version of the {\sc rebound}
code is available at~\url{https://github.com/hannorein/rebound}.



\bibliographystyle{mnras}
\bibliography{ref} 

\bsp	
\label{lastpage}
\end{document}